\documentclass{iau} 
\usepackage{graphicx}

\title[H$_{2}$O and SiO masers in Orion Source~I] %% give here short title %%
{ALMA observations of submillimeter H$_{2}$O and SiO lines in Orion Source~I}

\author[Tomoya Hirota et al.]   %% give here short author list %%
{Tomoya Hirota$^{1}$, 
Masahiro N. Machida$^{2}$,
Yuko Matsushita$^{2}$,
Kazuhito Motogi$^{3}$,
Naoko Matsumoto$^{1,3}$,
Mikyoung Kim$^{4,5}$,
Ross A. Burns$^{6}$, 
\and Mareki Honma$^{5}$}

\affiliation{
$^{1}$National Astronomical Observatory of Japan, Mitaka-shi, Tokyo 181-8588, Japan \\
email: {\tt tomoya.hirota@nao.ac.jp} \\[\affilskip]
$^{2}$Kyushu University, Fukuoka-shi 819-0395, Japan \\
$^{3}$Yamaguchi University, Yamaguchi-shi 753-8512, Japan \\
$^{4}$Korea Astronomy and Space Science Institute, Daejeon 305-348, Republic of Korea \\
$^{5}$National Astronomical Observatory of Japan, Oshu-shi, Iwate 023-0861, Japan \\
$^{6}$Joint Institute for VLBI ERIC, Postbus 2, 7990 AA Dwingeloo, The Netherlands \\
}

\pubyear{2017}
\volume{336}  %% insert here IAU Symposium No.
\jname{Astrophysical Masers: Unlocking the Mysteries of the Universe}
\editors{A. Tarchi, M.J. Reid \& P. Castangia, eds.}
\begin{document}

\maketitle

\begin{abstract}
We present observational results of the submillimeter H$_{2}$O and SiO lines toward a candidate high-mass young stellar object Orion Source~I using ALMA. 
The spatial structures of the high excitation lines at lower-state energies of $>$2500 K show compact structures consistent with the circumstellar disk and/or base of the northeast-southwest bipolar outflow with a 100 au scale. 
The highest excitation transition, the SiO ($v$=4) line at band 8, has the most compact structure. 
In contrast, lower-excitation transitions are more extended than 200 au tracing the outflow. 
Almost all the line show velocity gradients perpendicular to the outflow axis suggesting rotation motions of the circumstellar disk and outflow. 
While some of the detected lines show broad line profiles and spatially extended emission components indicative of thermal excitation, the strong H$_{2}$O lines at 321 GHz, 474 GHz, and 658 GHz with brightness temperatures of $>$1000 K show clear signatures of maser action. 
\end{abstract}

\firstsection 
\section{Introduction}

Orion Source~I is a candidate of a high-mass young stellar object (\cite[Menten \& Reid 1995]{Menten1995}) located in the nearest high-mass star-forming region Orion KL at a distance of $\sim$420 pc (\cite[Menten et al. 2007, Kim et al. 2008]{Menten2007, Kim2008}). 
It drives a low-velocity bipolar outflow along the northeast-southwest (NE-SW) direction with a 1000~au-scale in edge-on view (\cite[Greenhill et al. 2013]{Greenhill2013}). 
Source~I is known as one of the rare star-forming regions associated with the SiO masers (\cite[Menten \& Reid 1995]{Menten1995}). 
The vibrationally excited SiO masers trace a rotating outflow arising from a circumstellar disk with a $\sim$100~au scale (\cite[Kim et al. 2008, Matthews et al. 2010]{Kim2008, Matthews2010}). 
The rotation curve of the SiO masers implies an enclosed mass of (5$-$7)$M_{\odot}$ under the assumption of Keplerian rotation, which is also confirmed by recent observations with ALMA (\cite[Plambeck \& Wright 2016]{Plambeck2016}). 
However, the above mass estimate is significantly smaller than that from proper motion measurements with VLA, in which Source~I is proposed to be a 20~M$_{\odot}$ binary system formed by a dynamical encounter event 500 years ago (\cite[Goddi et al. 2011, Rodr\'iguez et al. 2017, Bally et al. 2017]{Goddi2011, Rodriguez2017, Bally2017}).  
Observations of Source~I at higher angular resolution would be crucial to investigate detailed physical and dynamical properties of Source~I. 
For this purpose, strong maser lines will be unique probes as they can reveal high density and temperature regions in close vicinity to newly born stars at higher resolution than weaker thermal molecular lines. 

\begin{table}
\begin{center}
\caption{Detected lines (ordered by the lower-state energy)}
\label{tab-lines}
{\scriptsize
\begin{tabular}{cllrccl}
%\hline
\hline
                                                   &
                                                   & 
\multicolumn{1}{c}{$\nu$}           &
\multicolumn{1}{c}{$E_{l}$}         &
\multicolumn{1}{c}{Beam size}  & 
\multicolumn{1}{c}{}  \\
\multicolumn{1}{c}{Molecule}   &
\multicolumn{1}{c}{Transition}  &
\multicolumn{1}{c}{(MHz)}          &
\multicolumn{1}{c}{(K)}               &
\multicolumn{1}{c}{(arcsec)}      &
\multicolumn{1}{c}{Reference}               \\
\hline
H$_{2}$O     & 5$_{3,3}$-4$_{4,0}$             & 474689   &   702    & 0.10"  & \\ 
H$_{2}$O     & 10$_{2,9}$-9$_{3,6}$           & 321226   & 1846    &  0.17" & \cite[Hirota et al. (2014)]{Hirota2014} \\
H$_{2}$O     & $v$=1, 1$_{1,0}$-1$_{0,1}$  & 658007   & 2329    &  0.26" & \cite[Hirota et al. (2016)]{Hirota2016} \\
H$_{2}$O     & $v$=1, 4$_{2,2}$-3$_{3,1}$  & 463171   & 2744    &  0.10" & \cite[Hirota et al. (2017)]{Hirota2017} \\
H$_{2}$O     & $v$=1, 5$_{2,3}$-6$_{1,6}$  & 336228   & 2939    &  0.17" & \cite[Hirota et al. (2014)]{Hirota2014}  \\
H$_{2}$O     & $v$=1, 5$_{5,0}$-6$_{4,3}$  & 232687   & 3451    &  0.18" & \cite[Hirota et al. (2012)]{Hirota2012}  \\
H$_{2}$O     & $v$=1, 7$_{4,4}$-6$_{5,1}$  & 498502   & 3673    &  0.09" \\
\hline
$^{29}$SiO   & 10-9                                      & 428684   &  93    & 0.08" & \\
SiO               & 11-10                                    & 477505   & 115    & 0.10" & \\
Si$^{18}$O   & 12-11                                    & 484056   & 128    & 0.09" & \cite[Hirota et al. (2017)]{Hirota2017} \\
$^{30}$SiO   & $v$=1, 11-10                        & 462757   & 1859    &  0.10" & \\
SiO               & $v$=1, 11-10                        & 474185   & 1883    &  0.10" & \\
$^{29}$SiO   & $v$=2, 11-10                        & 465014   & 3611    &  0.10" & \\
SiO               & $v$=2, 10-9                          & 428087   & 3614    &  0.08" & \\
SiO               & $v$=4, 11-10                        & 464245   & 7085    &  0.10" & \\
\hline
\end{tabular}
}
\end{center}
\end{table}
\begin{figure}[h]
\begin{center}
\includegraphics[width=4in]{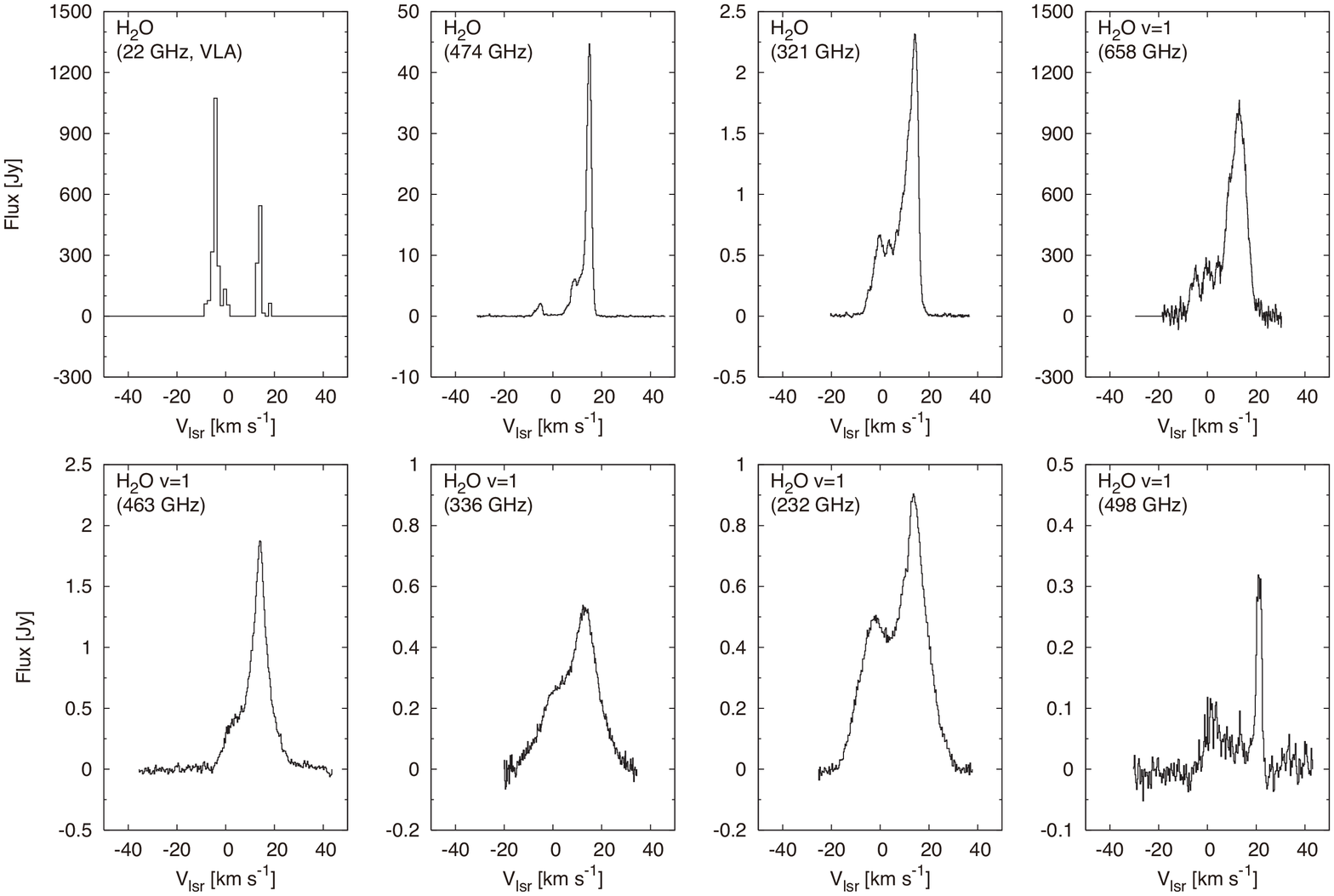} 
\caption{Spectra of the H$_{2}$O lines.  
For the 22~GHz (\cite[Gaume et al. 1998]{Gaume1998}) and 658~GHz lines, total flux densities are integrated over the primary beam. 
Flux densities for other lines are integrated over the synthesized beam size (Table \ref{tab-lines}) around peak positions of each line. }
\label{fig-h2o}
\end{center}
\end{figure}

\firstsection 
\section{Observations}

Observations of the submillimeter H$_{2}$O and SiO lines were carried out with ALMA in cycles 0, 1, and 2. 
We also compared the ALMA Science Verification data for Orion KL at band 6. 
Table \ref{tab-lines} summarizes the detected transitions. 
Details of observations and data analysis are described in previous papers (\cite[Hirota et al. 2012, 2014, 2016, 2017]{Hirota2012, Hirota2014, Hirota2016, Hirota2017}).  
   
\firstsection 
\section{Results}

\begin{figure}[t]
\begin{center}
\vspace*{5mm}
\includegraphics[width=4in]{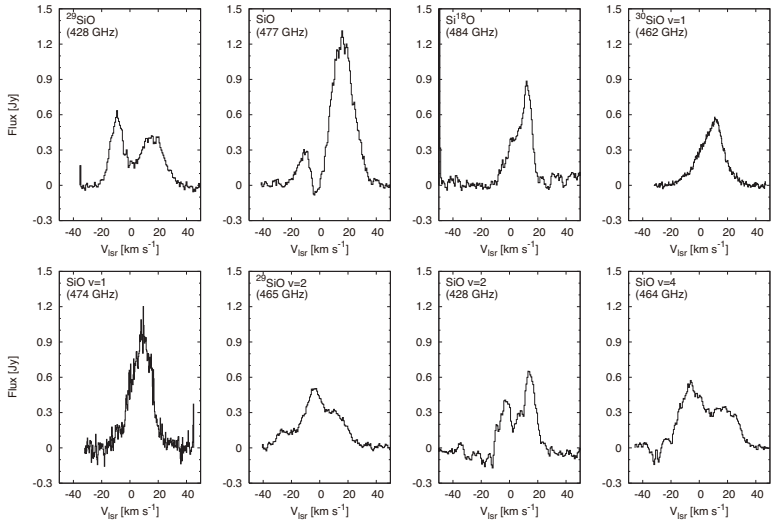} 
\caption{Spectra of the SiO lines observed toward the peak positions of each line. 
Flux densities are integrated over the beam size (Table \ref{tab-lines}).  
Note that the flux scales in all panels are common. }
\label{fig-sio}
\end{center}
\end{figure}

Figures \ref{fig-h2o} and \ref{fig-sio} show detected spectra of the H$_{2}$O and SiO lines, respectively. 
Intensive studies with single-dish telescopes have already detected several submillimeter H$_{2}$O lines in star-forming regions and late-type stars (\cite[Humphreys 2007]{Humphreys2007}) while the 463~GHz and 498~GHz lines are detected for the first time with our ALMA observations. 
All the spectra show broad line profiles and some of them have double peaked structures with the velocity width of $\sim$10-20~km~s$^{-1}$. 
These results are analogous to previously observed 22~GHz H$_{2}$O masers (Figure \ref{fig-h2o}; \cite[Gaume et al. 1998, Greenhill et al. 2013]{Gaume1998, Greenhill2013}) and 43~GHz SiO masers (\cite[Menten \& Reid 1995, Kim et al. 2008, Matthews et al. 2010]{Menten1995, Kim2008, Matthews2010}).

\begin{figure}[t]
\begin{center}
\includegraphics[width=4in]{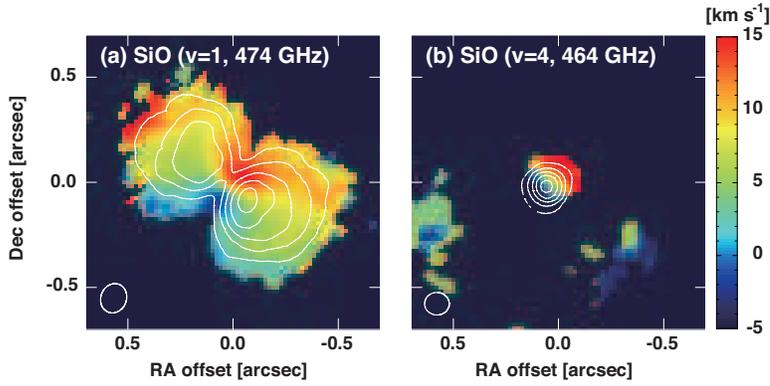} 
\caption{Moment 0 (cotour) maps of (a) SiO ($v$=1, 474~GHz) and (b) SiO ($v$=4, 464~GHz) lines superposed on their moment 1 maps.  
Contours represent 10, 30, 50, 70, and 90\% of the peak intensities of (a) 31~Jy~beam$^{-1}$~km~s$^{-1}$ and (b) 22~Jy~beam$^{-1}$~km~s$^{-1}$, respectively. 
}
\label{fig-maps}
\end{center}
\end{figure}

Recent ALMA observations have revealed that high excitation molecular lines trace a hot molecular gas disk and/or base of the NE-SW outflow in Source~I (\cite[Hirota et al. 2012, 2014, 2016, 2017, Plambeck \& Wright 2016]{Hirota2012, Hirota2014, Hirota2016, Hirota2017, Plambeck2016}).  
Figure \ref{fig-maps} shows examples of our new results  from the ALMA observations. 
We have found a clear trend that the higher excitation lines of H$_{2}$O and SiO at the lower state energy levels of $E_{l}>$2500~K could trace the compact disk or innermost region of the outflow with a 100~au scale, as can be seen in the vibrationally excited SiO masers at 43~GHz (\cite[Kim et al. 2008, Matthews et al. 2010]{Kim2008, Matthews2010}). 
The highest excitation transition we have detected is the $v$=4, $J$=11-10 transition of SiO ($E_{l}$=7085~K), which shows the most compact structure among the detected lines (Figure \ref{fig-maps}b). 
On the other hand, the lower excitation lines of $E_{l}<$2500~K trace the NE-SW outflow. 
The extended outflow structures can be traced even by the vibrationally excited $v$=1 transition of SiO (Figure \ref{fig-maps}a) with a size of $>$200~au. 
All the maps show velocity gradients perpendicular to the outflow axis, except SiO line at 477~GHz and H$_{2}$O line at 474~GHz, which are optically thick and are preferentially excited in the outer part of the outflow, respectively. 
These velocity structures, indicating rotation motion of the disk and outflow, provide clear evidence of magneto-centrifugal disk winds as a possible launching mechanism of the bipolar outflow from Source~I (\cite[Matthews et al. 2010, Greenhill et al. 2013, Hirota et al. 2017]{Matthews2010, Greenhill2013, Hirota2017}).

\firstsection 
\section{Discussions}

It is known that Source~I is associated with the strong H$_{2}$O maser at 22~GHz and SiO masers at 43~GHz (\cite[Menten \& Reid 1995, Gaume et al. 1998, Kim et al. 2008, Matthews et al. 2010, Greenhill et al. 2013]{Menten1995,Gaume1998, Kim2008, Matthews2010, Greenhill2013}). 
These maser emissions usually show spatially compact structures, higher brightness temperatures compared with gas kinetic temperatures, and narrow spike-like line profiles. 
Extremely high brightness temperatures of $>$1000~K of the 321~GHz, 474~GHz, and 658~GHz H$_{2}$O lines are clear signatures of maser emissions. 
It is also likely that a narrow spike-like spectral profile of the 498~GHz H$_{2}$O line (Figure \ref{fig-h2o}) could be an evidence of maser action. 
However, the other lines, in particular for SiO, show no clear signature of the above typical characteristics of masers. 
Flux density ratios of SiO isotopologues are different from those of isotope ratios (\cite[Tercero et al. 2011]{Tercero2011}) suggesting maser amplification and/or self-absorption in foreground gas. 
In order to access excitation mechanism of the observed lines, whether they are masers or thermal emissions, further maser pumping and radiative transfer models would be essential.

\firstsection
\section*{Acknowledgement}

This letter makes use of the following ALMA data: ADS/JAO.ALMA\#2011.0.00009.SV, ADS/JAO.ALMA\#2011.0.00199.S, and ADS/JAO.ALMA\#2013.1.00048.S. 
ALMA is a partnership of ESO (representing its member states), NSF (USA) and NINS (Japan), together with NRC (Canada), NSC and ASIAA (Taiwan), and KASI (Republic of Korea), in cooperation with the Republic of Chile. The Joint ALMA Observatory is operated by ESO, AUI/NRAO and NAOJ. 
TH is supported by the MEXT/JSPS KAKENHI Grant Numbers 24684011, 25108005, 15H03646, 16K05293, and 17K05398. 

\end{document}